\documentclass[12pt]{article}

\usepackage{enumitem}
\usepackage{graphicx}
\usepackage{float}
\usepackage{cite}
\usepackage{amsfonts}
\usepackage{amssymb}
\usepackage{amsmath}
\usepackage{mathtools}
\usepackage{xcolor}

\def\gtwid{\mathrel{\raise.3ex\hbox{$>$\kern-.75em\lower1ex\hbox{$\sim$}}}}
\def\ltwid{\mathrel{\raise.3ex\hbox{$<$\kern-.75em\lower1ex\hbox{$\sim$}}}}
\def\square{\kern1pt\vbox{\hrule height 1.2pt\hbox{\vrule width 1.2pt\hskip 3pt
   \vbox{\vskip 6pt}\hskip 3pt\vrule width 0.6pt}\hrule height 0.6pt}\kern1pt}

\begin{document}

\begin{titlepage}

\begin{flushright}
CCTP-2024-8 \\
UFIFT-QG-24-05
\end{flushright}

\vskip 0cm

\begin{center}
{\bf The Third Structure Function}
\end{center}

\vskip 0cm

\begin{center}
S. P. Miao$^{1\star}$, N. C. Tsamis$^{2\dagger}$, R. P. Woodard$^{3\ddagger}$
and B. Yesilyurt$^{3 *}$
\end{center}

\vskip 0cm

\begin{center}
\it{$^{1}$ Department of Physics, National Cheng Kung University, \\
No. 1 University Road, Tainan City 70101, TAIWAN}
\end{center}

\begin{center}
\it{$^{2}$ Institute of Theoretical Physics \& Computational Physics, \\
Department of Physics, University of Crete, \\
GR-700 03 Heraklion, HELLAS}
\end{center}

\begin{center}
\it{$^{3}$ Department of Physics, University of Florida,\\
Gainesville, FL 32611, UNITED STATES}
\end{center}

\vspace{0cm}

\begin{center}
ABSTRACT
\end{center}
We re-consider the graviton self-energy induced by a loop of massless, 
minimally coupled scalars on de Sitter background. On flat space background
it can be represented as a sum of two tensor differential operators acting
on scalar structure functions. On a general background these differential 
operators can be constructed from the linearized Ricci scalar and the 
linearized Weyl tensor. However, in cosmology one requires a third
contribution which we derive here.

\begin{flushleft}
\end{flushleft}

\vskip 0cm

\begin{flushleft}
$^{\star}$ e-mail: spmiao5@mail.ncku.edu.tw \\
$^{\dagger}$ e-mail: tsamis@physics.uoc.gr \\
$^{\ddagger}$ e-mail: woodard@phys.ufl.edu \\
$^{*}$ (Corresponding author) e-mail: b.yesilyurt@ufl.edu
\end{flushleft}

\end{titlepage}

\section{Prologue}

The background geometry of cosmology is characterized by a time-dependent
scale factor $a(t)$, and can be expressed either in terms of the co-moving
time $t$ or the conformal time $\eta$,
\begin{equation}
ds^2 = -dt^2 + a^2(t) d\vec{x} \!\cdot\! d\vec{x} = a^2 [-d\eta^2 +
d\vec{x} \!\cdot\! d\vec{x} ] \; . \label{background}
\end{equation}
We define the graviton field $h_{\mu\nu}(x)$ by perturbing the full metric
in conformal coordinates,
\begin{equation}
g_{\mu\nu} \equiv a^2 [\eta_{\mu\nu} + \kappa h_{\mu\nu}] \qquad , \qquad
\kappa^2 \equiv 16 \pi G \; . \label{graviton}
\end{equation}
The factor of $\kappa$ is included in the definition of $h_{\mu\nu}$ to cancel
the $\kappa^{-2}$ in the quadratic part of the Hilbert Lagrangian $\mathcal{L} 
= \kappa^{-2} R \sqrt{-g}$.
 
We call the 1PI (one-particle-irreducible) 2-point function of quantum gravity
the {\it graviton self-energy}, $-i [\mbox{}^{\mu\nu} \Sigma^{\rho\sigma}](x;x')$. 
It can be used to quantum-correct the linearized gravitational field equation,
\begin{equation}
\mathcal{D}^{\mu\nu\rho\sigma} \kappa h_{\rho\sigma}(x) 
= \tfrac12 \kappa^2 T^{\mu\nu}(x)
+ \int \!\! d^4x' \, [\mbox{}^{\mu\nu} \Sigma^{\rho\sigma}](x;x') 
\, \kappa h_{\rho\sigma}(x') 
\;\; . \label{Einstein}
\end{equation}

Here $\mathcal{D}^{\mu\nu\rho\sigma}$ is the Lichnerowicz operator, 
\begin{equation}
\kappa \tfrac{\delta S_{\rm Hilbert}}{\delta h_{\mu\nu}(x)} = a^2
\sqrt{-g} \, [-R^{\mu\nu} + \tfrac12 g^{\mu\nu} R] \equiv 
\mathcal{D}^{\mu\nu\rho\sigma} \kappa h_{\mu\nu} + O(\kappa^2 h^2) \; .
\label{Lichnerowicz}
\end{equation}
The symbol we call $T^{\mu\nu}$ in expression (\ref{Einstein}) is related
to the usual stress tensor,
\begin{equation}
T^{\mu\nu} \equiv -\tfrac{2}{\kappa} \tfrac{\delta S_{\rm matter}}{
\delta h_{\mu\nu}} = -a^2 \sqrt{-g} \, g^{\mu\rho} g^{\nu\sigma} \times -
\tfrac{2}{\sqrt{-g}} \tfrac{\delta S_{\rm matter}}{\delta g^{\rho\sigma}} 
\; . \label{stress}
\end{equation}
Equation (\ref{Einstein}) describes quantum corrections to free 
gravitational radiation, and to the gravitational response to specific matter 
sources. Although flat space background shows no change to gravitational 
radiation, and only small corrections to gravitational potentials 
\cite{Donoghue:1993eb,Donoghue:1994dn,Park:2010pj}, there can be dramatic 
effects on backgrounds such as de Sitter which generate copious particle 
production \cite{Wang:2015eaa,Park:2015kua,Tan:2021lza,Tan:2022xpn,
Miao:2024atw}.

The graviton self-energy is a 2nd rank, bi-tensor density on whatever 
background is under study, and the struggle to find a simple and useful way
of representing its tensor structure has been long and confusing 
\cite{Tsamis:1996qk,Park:2011ww,Leonard:2014zua,Tan:2021ibs}. An additional
complication is the distinction between matter loop contributions, which
obey a conservation relation at either point, and graviton loop corrections, 
which only obey the weaker relation of conservation at both points. We shall 
here consider the simpler case of matter loop corrections.

On flat, $D$-dimensional spacetime the tensor structure of matter contributions 
can be represented by spin zero and spin two combinations of the transverse 
projection operator, $\Pi^{\mu\nu} \equiv \partial^{\mu} \partial^{\nu} - 
\eta^{\mu\nu} \partial^2$ \cite{Park:2010pj},
\begin{equation}
-i[\mbox{}^{\mu\nu} \Sigma^{\rho\sigma}_{\rm flat}](x;x') = \Pi^{\mu\nu} 
\Pi^{\rho\sigma} F_0(x-x') + \Bigl[ \Pi^{\mu (\rho} \Pi^{\sigma) \nu} 
- \tfrac1{D-1} \Pi^{\mu\nu} \Pi^{\rho\sigma}\Bigr] F_2(x - x') \; , 
\label{flatrep}
\end{equation}
where parenthesized indices are symmetrized. For a massless, minimally 
coupled scalar in $D=4$ the renormalized, Schwinger-Keldysh\footnote{
We use the in-in, Schwinger-Keldysh formalism because the 
graviton self-energy of the usual, in-out formalism is neither real, nor 
causal. The Schwinger-Keldysh formalism gives true expectation values
\cite{Schwinger:1960qe,Mahanthappa:1962ex,Bakshi:1962dv,Bakshi:1963bn,
Keldysh:1964ud}, unlike the in-out matrix elements of the usual formalism.
The effective field equations of the Schinger-Keldysh formalism are real
and causal \cite{Chou:1984es,Jordan:1986ug,Calzetta:1986ey}, as is manifest 
in expressions (\ref{F0flat}-\ref{F2flat}).} structure functions are 
\cite{Campos:1993ug,Martin:2000dda,Ford:2004wc},
\begin{eqnarray}
i F_0(\Delta x) &\!\!\! = \!\!\!& -\frac{\kappa^2}{2^{10} \!\cdot\! 3^2 
\!\cdot\! \pi^3} \!\times\! \partial^4 \Bigl\{ \theta(\Delta t \!-\! \Delta r) 
\Bigl(\ln[\mu^2 (\Delta t^2 \!-\! \Delta r^2)] \!-\! 1\Bigr) \Bigr\} , 
\qquad \label{F0flat} \\
i F_2(\Delta x) &\!\!\! = \!\!\!& -\frac{\kappa^2}{2^{11} \!\cdot\! 3
\!\cdot\! 5 \!\cdot\!\pi^3} 
\!\times\! \partial^4 \Bigl\{ \theta(\Delta t \!-\! \Delta r) \Bigl(
\ln[\mu^2 (\Delta t^2 \!-\! \Delta r^2)] \!-\! 1\Bigr) \Bigr\} , \qquad
\label{F2flat}
\end{eqnarray}
where $\Delta t \equiv t - t'$, $\Delta r \equiv \Vert \vec{x} - \vec{x}'
\Vert$ and $\mu$ is the renormalization scale.
 
We seek a similar representation on de Sitter background in conformal 
coordinates, for which the scale factor is,\footnote{
The relation (\ref{deSitter}) between co-moving time $t$ and 
conformal time $\eta$ comes from $d\eta = dt/a(t)$, with the integration
constant fixed by requiring that $t \rightarrow \infty$ corresponds to
$\eta \rightarrow 0^{-}$.}
\begin{equation}
a(t) = e^{H t} = -\tfrac1{H \eta} \; , \label{deSitter}
\end{equation}
where the Hubble constant $H$ is related to the cosmological constant
$\Lambda \equiv (D-1) H^2$. We will continue to employ the flat space 
convention that indices are raised and lowered using the Minkowski metric
($\partial^{\mu} \equiv \eta^{\mu\nu} \partial_{\nu}$). The appropriate
de Sitter generalization of transversality is,
\begin{equation}
\mathcal{W}^{\mu}_{~\alpha\beta} \!\times\! -i [\mbox{}^{\alpha\beta}
\Sigma^{\rho\sigma}](x;x') = 0 = {\mathcal{W}}'^{\rho}_{~\gamma\delta}
\!\times\! -i [\mbox{}^{\mu\nu} \Sigma^{\gamma\delta}](x;x') \; ,
\label{conservation}
\end{equation}
where the Ward operator is \cite{Leonard:2014zua},
\begin{equation}
\mathcal{W}^{\mu}_{~\alpha\beta} \equiv \delta^{\mu}_{~(\alpha} \partial_{\beta)} 
+ a H \delta^{\mu}_{~0} \eta_{\alpha\beta} \; . \label{Wardop}
\end{equation}
In (\ref{conservation}) and throughout the paper, having a prime superscript on any differential operator means that the operator depends on primed coordinates, and thus, acts on primed coordinates.
\\
\indent The spin zero tensor in relation (\ref{flatrep}) can be generalized to de 
Sitter using a differential operator $\mathcal{F}^{\mu\nu}$,
\begin{equation}
\Pi^{\mu\nu} \Pi^{\rho\sigma} \longrightarrow \mathcal{F}^{\mu\nu} \!\times\!
{\mathcal{F}'}^{\rho\sigma} \; . \label{dS0}
\end{equation}
One defines this by first expanding the Ricci tensor to first order in the
graviton field $g_{\mu\nu} \equiv a^2 [\eta_{\mu\nu} + \kappa h_{\mu\nu}]$,
\begin{equation}
R - D \Lambda \equiv \tfrac1{a^2} \!\times\! \overline{\mathcal{F}}^{\mu\nu}
\!\times\! \kappa h_{\mu\nu} + O(h^2) \; . \label{Fbar}
\end{equation}
The operator $\mathcal{F}^{\mu\nu}$ is obtained by partially integrating 
$\overline{\mathcal{F}}^{\mu\nu}$,
\begin{equation}
\int \!\! d^Dx \, f_1(x) \!\times\! \overline{\mathcal{F}}^{\mu\nu} f_2(x)
\equiv \int \!\! d^Dx \, \mathcal{F}^{\mu\nu} f_1(x) \!\times\! f_2(x) \; .
\label{Fdef}
\end{equation}
The final result is an obvious de Sitter generalization of $\Pi^{\mu\nu}$, 
\begin{eqnarray}
\lefteqn{\mathcal{F}^{\mu\nu} = \partial^{\mu} \partial^{\nu} - \eta^{\mu\nu}
\partial^2 } \nonumber \\
& & \hspace{1.5cm} + (D\!-\!1) a H \Bigl[2 \delta^{(\mu}_{~~0} \partial^{\nu)}
+ (D\!-\!2) a H \delta^{\mu}_{~0} \delta^{\nu}_{~0} - \eta^{\mu\nu}
\partial_0 - \eta^{\mu\nu} a H \Bigr] . \label{Fexplicit} \qquad
\end{eqnarray}
It is simple to check that $\mathcal{W}^{\mu}_{~\alpha\beta} \times 
\mathcal{F}^{\alpha\beta} = 0$.

The spin two tensor in relation (\ref{flatrep}) can be generalized to
de Sitter using a differential operator $\mathcal{C}_{\alpha
\beta\gamma\delta}^{~~~~~\mu\nu}$,
\begin{equation}
\Pi^{\mu (\rho} \Pi^{\sigma) \nu} - \tfrac1{D-1} \Pi^{\mu\nu} 
\Pi^{\rho\sigma} \longrightarrow \Bigl( \frac{D\!-\!2}{D\!-\!3}
\Bigr) \!\times\! \mathcal{C}_{\alpha\beta\gamma\delta}^{~~~~~\mu\nu} 
\!\times\! {\mathcal{C}'}^{\alpha\beta\gamma\delta\rho\sigma} \; . 
\label{dS2}
\end{equation}
One defines $\mathcal{C}_{\alpha\beta\gamma\delta}^{~~~~~\mu\nu}$ by 
expanding the Weyl tensor to first order in the graviton field,
\begin{equation}
\eta_{\alpha\epsilon} C^{\epsilon}_{~\beta\gamma\delta} \equiv
\mathcal{C}_{\alpha\beta\gamma\delta}^{~~~~~\mu\nu} \!\times\! \kappa
h_{\mu\nu} + O(h^2) \; . \label{Cdef}
\end{equation}
It is not necessary to partially integrate because each term in 
$\mathcal{C}_{\alpha\beta\gamma\delta}^{~~~~~\mu\nu}$ contains two 
derivatives, with no scale factors. An explicit form can be found 
in published works \cite{Park:2011ww,Leonard:2014zua}. Like 
$\mathcal{F}^{\mu\nu}$, it is annihilated by the Ward operator,
\begin{equation}
\mathcal{W}^{\lambda}_{~\mu\nu} \!\times\! \mathcal{C}_{\alpha\beta
\gamma\delta}^{~~~~~\mu\nu} = 0 \; . \label{WC0}
\end{equation}

It turns out that representing the graviton self-energy on cosmological
backgrounds such as (\ref{deSitter}) requires a third structure function
\cite{Leonard:2014zua}, and finding a simple representation for this has
proven to be difficult. Because relation (\ref{WC0}) pertains for any 
choices of the first four indices of $\mathcal{C}_{\alpha\beta\gamma
\delta}^{~~~~~\mu\nu}$, conservation (\ref{conservation}) would apply 
were we to contract any constant index factor into the product,
\begin{equation}
\mathcal{C}_{\alpha\beta\gamma\delta}^{~~~~~\mu\nu} \!\times\!
{\mathcal{C}'}_{\kappa\lambda\theta\phi}^{~~~~~\rho\sigma} \; .
\end{equation}
The particular choice giving rise to (\ref{dS2}) is the product of four
Minkowski metrics,
\begin{equation}
\eta^{\alpha\kappa} \!\times\! \eta^{\beta\lambda} \!\times\!
\eta^{\gamma\theta} \!\times\! \eta^{\delta\phi} \; . \label{choice0}
\end{equation}
So a natural thought is to define a distinct projector by replacing
(\ref{choice0}) with another constant index factor, such as the product of four 
spatial metrics,
\begin{equation}
\overline{\eta}^{\mu\nu} \equiv \eta^{\mu\nu} + \delta^{\mu}_{~0}
\delta^{\nu}_{~0} \; . \label{spaceeta}
\end{equation}
However, when this is done the structure functions become quite 
complicated \cite{Leonard:2014zua} and difficult to use \cite{Park:2015kua}.
Further, assuming this form for the third projector seems ill-considered 
because, like the de Sitter generalizations (\ref{dS0}) and (\ref{dS2}), it
is a 4th order differential operator. Because the need for the third 
projector is tied to the nonzero Hubble parameter of cosmological backgrounds,
it would make more sense to seek a 2nd order differential operator,
with two factors of $H$ in place of the two missing derivatives. 

The number of structure functions that should be necessary
to characterize the graviton self-energy on a general cosmological background 
can be easily derived from symmetry arguments. First, note that the cosmological
symmetries of homogeneity and isotropy dictate that it can be expressed as a
linear combination of the 21 differential operators of Table~\ref{Tbasis}
acting on coefficient functions $F^i(x;x)$,
\begin{equation}
-i [\mbox{}^{\mu\nu} \Sigma^{\rho\sigma}](x;x') = \sum_{i=1}^{21}
[\mbox{}^{\mu\nu} D_i^{\rho\sigma}] \!\times\! F^i(x;x') \; . \label{genform}
\end{equation}
Now use reflection invariance (which follows from the graviton self-energy 
representing the expectation value of a time-ordered product of two fields),
\begin{equation}
-i [\mbox{}^{\mu\nu} \Sigma^{\rho\sigma}](x;x') = -i [\mbox{}^{\rho\sigma} 
\Sigma^{\mu\nu}](x';x) \ . \label{refinvariance}
\end{equation}
This implies 7 relations between the coefficient functions \cite{Tan:2021ibs},
\begin{eqnarray}
i=3,7,16 & \Longrightarrow & F^i(x;x') = +F^{i+1}(x';x) \; , \label{refl1} \\
i=5,10,14,19 & \Longrightarrow & F^i(x;x') = -F^{i+1}(x';x) \; . \label{refl2}
\end{eqnarray}
The remaining functions obey $F^i(x;x') = F^i(x';x)$. Now impose the requirement
of conservation (\ref{conservation}), which turns out to give ten conditions 
\cite{Leonard:2014zua} that we will not list here. It follows that the graviton 
self-energy can require up to four independent functions,
\begin{equation}
21 - 7 - 10 = 4 \; .
\end{equation}
The first two of these structure functions are the de Sitter descendants of 
$F_0(\Delta x)$ and $F_2(\Delta x)$ in expression (\ref{flatrep}). 
A proposal for the remaining two was made involving automatically
conserved, 4th order tensor differential operators acting on scalar structure 
functions, similar to $F_0$ and $F_2$ \cite{Leonard:2014zua}.\footnote{
It turns out that one of these extra structure functions 
vanishes for the contribution from a massless, minimally coupled scalar 
as a consequence of de Sitter invariance\cite{Leonard:2014zua}.} However, 
this representation results in very complicated structure functions, and 
its application to the original calculation of a loop of massless, 
minimally coupled scalars \cite{Park:2011ww,Leonard:2014zua} was invalid 
owing to absence of a finite renormalization of the cosmological constant 
which must be done to make the graviton self-energy fully conserved 
\cite{Tsamis:2023fri}. The point of this work is to identify a 3rd (and 
possibly 4th) contribution to the graviton self-energy as $H^2$ times some 
second order differential operator, acting on a simple structure function. 

In Section 2 we begin with a recent computation
of the massless, minimally coupled scalar loop contribution to $-i[\mbox{}^{\mu\nu}
\Sigma^{\rho\sigma}](x;x')$ and extract the terms coming from the spin 0 and
spin 2 projectors (\ref{dS0}) and (\ref{dS2}). In Section 3 we promote the
Lichnerowicz operator into a bi-tensor. In Section 4 it is made the basis for
expressing the 3rd contribution. Our conclusions comprise Section 5.

\section{The First Two Structure Functions}

The purpose of this section is to express the previously computed
result for the (massless, minimally coupled scalar contribution to
the) graviton self-energy \cite{Miao:2024atw}as a sum of  
contributions involving $\mathcal{F}^{\mu\nu} \times {\mathcal{F}'}^{
\rho\sigma}$ and $\mathcal{C}_{\alpha\beta\gamma\delta}^{~~~~~\mu\nu} 
\times {\mathcal{C}'}^{\alpha\beta\gamma\delta\rho\sigma}$, acting on 
structure functions, plus a separately conserved residual. It is this 
residual which will eventually be expressed as the 3rd structure function. 
We begin by introducing some useful notation. Then the previous result for 
$[\mbox{}^{\mu\nu} \Sigma^{\rho\sigma}](x;x')$ is given. The section closes 
by extracting the parts associated with the (\ref{dS0}) and (\ref{dS2}) 
projectors to leave the residual we seek to study.

\begin{table}[H]
\setlength{\tabcolsep}{8pt}
\def\arraystretch{1.5}
\centering
\begin{tabular}{|@{\hskip 1mm }c@{\hskip 1mm }||c||c|c||c|c|}
\hline
$i$ & $[\mbox{}^{\mu\nu} D^{\rho\sigma}_i]$ & $i$ & 
$[\mbox{}^{\mu\nu} D^{\rho\sigma}_i]$
& $i$ & $[\mbox{}^{\mu\nu} D^{\rho\sigma}_i]$ \\
\hline\hline
1 & $\eta^{\mu\nu} \eta^{\rho\sigma}$ & 8 & 
$\partial^{\mu} \partial^{\nu} \eta^{\rho\sigma}$ & 15 & 
$\delta^{(\mu}_{~~0} \partial^{\nu)} 
\delta^{\rho}_{~0} \delta^{\sigma}_{~0}$ \\
\hline
2 & $\eta^{\mu (\rho} \eta^{\sigma) \nu}$ & 9 & 
$\delta^{(\mu}_{~~0} \eta^{\nu) (\rho}
\delta^{\sigma)}_{~~0}$ & 16 & 
$\delta^{\mu}_{~0} \delta^{\nu}_{~0} \partial^{\rho}
\partial^{\sigma}$ \\
\hline
3 & $\eta^{\mu\nu} \delta^{\rho}_{~0} \delta^{\sigma}_{~0}$ & 10 & 
$\delta^{(\mu}_{~~0} \eta^{\nu) (\rho} \partial^{\sigma)}$ & 17 & 
$\partial^{\mu} \partial^{\nu} \delta^{\rho}_{~0}
\delta^{\sigma}_{~0}$ \\
\hline
4 & $\delta^{\mu}_{~0} \delta^{\nu}_{~0} \eta^{\rho\sigma}$ & 11 & 
$\partial^{(\mu} \eta^{\nu) (\rho} \delta^{\sigma)}_{~~0}$ & 18 & 
$\delta^{(\mu}_{~~0} \partial^{\nu)}
\delta^{(\rho}_{~~0} \partial^{\sigma)}$ \\
\hline
5 & $\eta^{\mu\nu} \delta^{(\rho}_{~~0} \partial^{\sigma)}$ & 12 & 
$\partial^{(\mu} \eta^{\nu)(\rho} \partial^{\sigma)}$ & 19 & 
$\delta^{(\mu}_{~~0} \partial^{\nu)}
\partial^{\rho} \partial^{\sigma}$ \\
\hline
6 & $\delta^{(\mu}_{~~0} \partial^{\nu)} \eta^{\rho\sigma}$ & 13 & 
$\delta^{\mu}_{~0} \delta^{\nu}_{~0} \delta^{\rho}_{~0} 
\delta^{\sigma}_{~0}$ & 20 & 
$\partial^{\mu} \partial^{\nu} 
\delta^{(\rho}_{~~0} \partial^{\sigma)}$ \\
\hline
7 & $\eta^{\mu\nu} \partial^{\rho} \partial^{\sigma}$ & 14 & 
$\delta^{\mu}_{~0} \delta^{\nu}_{~0} \delta^{(\rho}_{~~0} 
\partial^{\sigma)}$ & 21 & 
$\partial^{\mu} \partial^{\nu} \partial^{\rho} \partial^{\sigma}$ \\
\hline
\end{tabular}
\caption{\footnotesize 
The 21 basis differential operators.}
\label{Tbasis}
\end{table} 

The massless, minimally coupled scalar contribution to the graviton 
self-energy on de Sitter background was first computed more than a decade 
ago \cite{Park:2011ww}.\footnote{The computation was made
in Bunch-Davies vacuum \cite{Chernikov:1968zm,Schomblond:1976xc,
Bunch:1978yq}. This is the usual choice for computations on de Sitter 
background and corresponds to a free field theory in which the universe 
was empty in the distant past. When interactions are present one must
generally make perturbative corrections to the wave functional of the
initial state \cite{Kahya:2009sz}.} However, it was not then realized 
that a finite renormalization of the cosmological constant is required to 
prevent the appearance of a delta function obstruction to conservation 
\cite{Tsamis:2023fri}. This mistake was compounded by choosing to express
the structure function in terms of projectors which are annihilated by the
Ward operator \cite{Leonard:2014zua}. The two errors might have canceled out
but it turns out they do not \cite{Miao:2024atw}. In order to check, and to
facilitate a subsequent resummation \cite{Miao:2024nsz}, the result was 
recomputed in the safest way possible (by which we mean
using the basic diagrams, without assuming they can be expressed in terms
of conserved tensor differential operators acting on structure functions),
expressing it as a sum of the 21 (non-conserved) differential operators 
given in Table~\ref{Tbasis}, whose form is dictated by the homogeneity and 
isotropy of cosmology.

\begin{table}[H]
\setlength{\tabcolsep}{8pt}
\def\arraystretch{1.25}
\centering
\begin{tabular}{|@{\hskip 1mm }c@{\hskip 1mm }||c||c|}
\hline
$i$ & $T^i(a,a',\partial)$ & $T_A^i(a,a',\partial)$ \\
\hline\hline
1 & $-a a' \ln(a a') [\partial^2 + 2 a a' H^2]$ & 
$-\frac14 a^2 {a'}^2 H^2 \partial^2 \partial_0^2$ \\
& $-3 a a' \partial^2 - a a' \partial_0^2 - 6 a^2 {a'}^2 H^2$ & $$ \\
\hline
2 & $a a' \ln(a a') [\partial^2 + 2 a a' H^2]$ & $0$ \\
& $+ a a' \partial^2 + 3 a^2 {a'}^2 H^2$ & \\
\hline
3 & $2 a^3 a' H^2 [\ln(a a') + \frac92]$ & $0$ \\
\hline
4 & $2 a {a'}^3 H^2 [\ln(a a') + \frac92]$ & $0$ \\
\hline
5 & $-2 a^2 a' H [\ln(a a') + 4]$ & 
$\frac12 a^2 {a'}^2 H^2 \partial_0 \partial^2$ \\
\hline
6 & $2 a {a'}^2 H [\ln(a a') + 4]$ & 
$\frac12 a^2 {a'}^2 H^2 \partial_0 \partial^2$ \\
\hline
7 & $a a' [\ln(a a') + 3]$ & 
$- \frac{1}{2} a^2 a' H \partial_0 \partial^2 
+ \frac{1}{2} a^2 {a'}^2 H^2 \partial^2$ \\
\hline
8 & $a a' [\ln(a a') + 3]$ & 
$\frac{1}{2} a {a'}^2 H \partial_0 \partial^2 
+ \frac{1}{2} a^2 {a'}^2 H^2 \partial^2$ \\
\hline
9 & $2 a^2 {a'}^2 H^2 [\ln(a a') + 1]$ & $0$ \\
\hline
10 & $-2 a {a'}^2 H [\ln(a a') + 2]$ & $0$ \\
\hline 
11 & $2 a^2 a' H [\ln(a a') + 2]$ & $0$ \\
\hline
12 & $-2 a a' [\ln(a a') + 1]$ & $-a^2 {a'}^2 H^2 \partial^2$ \\
\hline
18 & $0$ & $-3a^2 {a'}^2 H^2 \partial^2$ \\
\hline
19 & $0$ & $a^2 a' H \partial^2$ \\
\hline
20 & $0$ & $-a {a'}^2 H \partial^2$ \\
\hline
21 & $0$ & $\frac{1}{2} a a' \partial^2 - a^2 {a'}^2 H^2$ \\
\hline
\end{tabular}
\caption{\footnotesize Coefficients $T^i(a,a',\partial)$ and 
$T^i_{A}(a,a',\partial)$ which appear in expression (\ref{Sigma1}).}
\label{TTA}
\end{table}

Finite remainders from the $a^{D-4} R^2$ and $a^{D-4} C^2$ counterterms give 
rise to factors of $\ln(a a') \delta^4(\Delta x)$ acted on by the de Sitter 
spin 0 and spin 2 projectors (\ref{dS0}) and (\ref{dS2}). The remaining terms 
involve $\delta^4(\Delta x)$ and two other functions,
\begin{equation}
f_A(\Delta x) \equiv \theta(\Delta \eta \!-\! \Delta r) \quad , \quad 
f_B(\Delta x) \equiv \partial^4 \Bigl\{ \theta(\Delta \eta \!-\! \Delta r) 
\Bigl( \ln[ \mu^2 (\Delta \eta^2 \!-\! \Delta r^2)] \!-\! 1\Bigr\} \; , 
\label{fAfB}
\end{equation}
where $\Delta \eta \equiv \eta - \eta'$ and $\Delta r \equiv \Vert \vec{x}
- \vec{x}' \Vert$. Coefficient functions of the two scale factors and the
derivative operator (given in Tables~\ref{TTA} and \ref{TB}) times the basis
operators of Table~\ref{Tbasis} act on these three functions \cite{Miao:2024atw},
\begin{eqnarray}
\lefteqn{ [\mbox{}^{\mu\nu} \Sigma^{\rho\sigma}](x;x') = \frac{-\kappa^2}{
2^7 \!\cdot\! 3 \!\cdot\! \pi^2} \Bigl[ \tfrac13 \mathcal{F}^{\mu\nu} 
\!\!\times\! {\mathcal{F}'}^{\rho\sigma} \!\!+\! \tfrac15 \mathcal{C}_{\alpha
\beta\gamma\delta}^{~~~~~\mu\nu} \!\!\times\! {\mathcal{C}'}^{\alpha\beta\gamma
\delta\rho\sigma} \Bigr] \Bigl[ \ln(a a')  \delta^4(\Delta x) \Bigr]
} \nonumber \\
& & \hspace{-0.5cm} + \frac{\kappa^2 H^2}{2^7 \!\cdot\! 3 \!\cdot\! \pi^3} 
\sum_{i=1}^{21} \Bigl\{-2\pi T^i(a,a',\partial) [\mbox{}^{\mu\nu} 
D^{\rho\sigma}_{i}] \delta^4(\Delta x) + T^i_{A}(a,a'\!,\partial) 
[\mbox{}^{\mu\nu} D^{\rho\sigma}_{i}] f_A (\Delta x) \nonumber \\
& & \hspace{6.9cm} + T^i_{B}(a,a'\!,\partial) [\mbox{}^{\mu\nu} 
D^{\rho\sigma}_{i}] f_B(\Delta x) \Bigr\} . \label{Sigma1} \qquad
\end{eqnarray}
\begin{table}[H]
\setlength{\tabcolsep}{8pt}
\def\arraystretch{1.5}
\centering
\begin{tabular}{|@{\hskip 1mm }c@{\hskip 1mm }||c||c||c|}
\hline
$i$ & $T_B^i(a,a',\partial)$ & $i$ & $T_B^i(a,a',\partial)$ \\
\hline\hline 
1  & $-\frac{3 \partial^4}{80 H^2} + \frac{a a'\partial^2}{8} 
   + \frac{a a' \partial_0^2}{8}$ & 
12 & $\frac{\partial^2}{40 H^2} + \frac{a a'}{2}$ \\
\hline
2 & $-\frac{\partial^4}{80 H^2} - \frac{a a' \partial^2}{4} 
  - \frac{a^2 {a'}^2 H^2}{2}$ & 
13 & $-\frac{3 a^2 {a'}^2 H^2}{2}$ \\
\hline
3 & $\frac{{a'}^2 \partial^2}{4} + \frac{3 a {a'}^2 H \partial_0}{4}
  + \frac{a^2 {a'}^2 H^2}{2}$ & 
14 & $\frac{3 a^2 a' H}{2}$ \\
\hline
4 & $\frac{a^2 \partial^2}{4} - \frac{3 a^2 a' H \partial_0}{4}
  + \frac{a^2 {a'}^2 H^2}{2}$ & 
15 & $-\frac{3 a {a'}^2 H}{2}$ \\
\hline
5 & $-\frac{a' \partial^2}{4 H} - \frac{3 a a' \partial_0}{4} 
  + \frac{a {a'}^2 H}{4}$ & 
16 & $-\frac{a^2}{4}$ \\
\hline
6 & $\frac{a \partial^2}{4 H} - \frac{3 a a' \partial_0}{4} 
  - \frac{a^2 a' H}{4}$ & 
17 & $-\frac{{a'}^2}{4}$ \\
\hline
7 & $\frac{3 \partial^2}{80 H^2} + \frac{a \partial_0}{8 H} 
  - \frac{a^2}{8}$ &
18 & $\frac{3 a a'}{2}$ \\
\hline
8 & $\frac{3 \partial^2}{80 H^2} - \frac{a' \partial_0}{8 H} 
  - \frac{{a'}^2}{8}$ &
19 & $-\frac{a}{4 H}$ \\
\hline
9 & $-\frac{3 a^2 {a'}^2 H^2}{2}$ &
20 & $\frac{a'}{4 H}$ \\
\hline
10 & $\frac{a^2 a' H}{2}$ &
21 & $-\frac1{20 H^2}$ \\
\hline
11 & $-\frac{a {a'}^2 H}{2}$ & & $$ \\ 
\hline
\end{tabular}
\caption{\footnotesize Coefficient functions $T^i_{B}(a,a',\partial)$ 
which appear in expression (\ref{Sigma1}).}
\label{TB}
\end{table}
\noindent Note that we do not need to distinguish $\partial'_{\mu} 
\rightarrow -\partial_{\mu}$ when acting on functions of $\Delta x^{\mu} 
= x^{\mu} - {x'}^{\mu}$. Note also that the three functions obey,
\begin{equation}
\partial^4 f_A = 8\pi \delta^4(\Delta x) \;\; , \;\;
[\Delta \eta \partial^2 + 2 \partial_0] f_A = 0 \;\; , \;\; 
\Delta \eta f_B = -2 \partial_0 \partial^2 f_A \; . 
\label{relations}
\end{equation}

\begin{table}[H]
\setlength{\tabcolsep}{8pt}
\def\arraystretch{1.5}
\centering
\begin{tabular}{|@{\hskip 1mm }c@{\hskip 1mm }||c|c|c|}
\hline
$i$ & $\Delta T^i(a,a',\partial)$ & $\Delta T^i_{A}(a,a',\partial)$ & 
$\Delta T^i_{B}(a,a',\partial)$ \\
\hline\hline
1 & $\!\!\!-a a' \ln(a a') [\partial^2 \!+\! 2 a a' H^2]\!\!\!$ & $0$ & 
$\!\!\!\tfrac14 a a' \partial^2 \!+\! \tfrac12 a^2 {a'}^2 H^2\!\!\!$ \\
  & $\!\!\!-3 a a' \partial^2 \!-\! 7 a^2 {a'}^2 H^2\!\!\!$ & $$ & $$ \\
\hline
2 & $\!\!\!a a' \ln(a a') [\partial^2 \!+\! 2 a a' H^2]\!\!\!$ & $0$ & 
$\!\!\!-\tfrac14 a a' \partial^2 \!-\! \tfrac12 a^2 {a'}^2 H^2\!\!\!$ \\
  & $\!\!\!+ a a' \partial^2 \!+\! 3 a^2 {a'}^2 H^2\!\!\!$ & $$ & $$ \\
\hline
3 & $\!\!\!a^3 a' H^2 [2 \ln(a a') \!+\! 9]\!\!\!$ & $0$ & $\!\!\!-\tfrac14 a^2 
{a'}^2 H^2\!\!\!$ \\
\hline
4 & $\!\!\!a {a'}^3 H^2 [2 \ln(a a') \!+\! 9]\!\!\!$ & $0$ & $\!\!\!-\tfrac14 a^2 
{a'}^2 H^2\!\!\!$ \\
\hline
5 & $\!\!\!-a^2 a' H [2 \ln(a a') \!+\! 8]\!\!\!$ & $\!\!\!\tfrac12 a^2 {a'}^2
H^2 \partial_0 \partial^2\!\!\!$ & $\!\!\!\tfrac14 a {a'}^2 H \!+\! \tfrac14
a^2 a' H\!\!\!$ \\
\hline
6 & $\!\!\!a {a'}^2 H [2 \ln(a a') \!+\! 8]\!\!\!$ & $\!\!\!\tfrac12 a^2 {a'}^2
H^2 \partial_0 \partial^2\!\!\!$ & $\!\!\!-\tfrac14 a {a'}^2 H \!-\! \tfrac14
a^2 a' H\!\!\!$ \\
\hline
7 & $\!\!\!a a'[\ln(a a') \!+\! 3]\!\!\!$ & $\!\!\!\tfrac12 a^2 {a'}^2 H^2 
\partial^2\!\!\!$ & $\!\!\!-\tfrac14 a a'$ \\
\hline
8 & $\!\!\!a a'[\ln(a a') \!+\! 3]\!\!\!$ & $\!\!\!\tfrac12 a^2 {a'}^2 H^2 
\partial^2\!\!\!$ & $\!\!\!-\tfrac14 a a'\!\!\!$ \\
\hline
9 & $\!\!\!a^2 {a'}^2 H^2 [2 \ln(a a') \!+\! 2]\!\!\!$ & $0$ & $\!\!\!-\tfrac32
a^2 {a'}^2 H^2\!\!\!$ \\
\hline
10 & $\!\!\!-a {a'}^2 H [2 \ln(a a') \!+\! 4]\!\!\!$ & $0$ & $\!\!\!\tfrac12
a^2 a' H\!\!\!$ \\
\hline
11 & $\!\!\!a^2 a' H [2 \ln(a a') \!+\! 4]\!\!\!$ & $0$ & $\!\!\!-\tfrac12
a {a'}^2 H\!\!\!$ \\
\hline
12 & $\!\!\!-a a' [2 \ln(a a') \!+\! 2]\!\!\!$ & $\!\!\!-a^2 {a'}^2 H^2 
\partial^2\!\!\!$ & $\!\!\!\tfrac12 a a'\!\!\!$ \\
\hline
18 & $0$ & $\!\!\!-3 a^2 {a'}^2 H^2 \partial^2\!\!\!$ & $0$ \\
\hline
19 & $0$ & $\!\!\!a^2 a' H \partial^2\!\!\!$ & $0$ \\
\hline
20 & $0$ & $\!\!\!-a {a'}^2 H \partial^2\!\!\!$ & $0$ \\
\hline
21 & $0$ & $\!\!\!\tfrac12 a a' \partial^2 \!-\! a^2 {a'}^2 H^2\!\!\!$ & $0$ \\
\hline
\end{tabular}
\caption{\footnotesize Tabulation of the residual coefficient functions
which appear in expression (\ref{Sigma2}).}
\label{Residual}
\end{table}

Expression (\ref{Sigma1}) for the graviton self-energy was employed because 
it is safe in the sense that one can still represent the residual contributions as a sum of 21 differential operators given in Table \ref{Tbasis}, however, it can easily be simplified. The key is recognizing the 
function $f_B(\Delta x)$ in the flat space structure functions (\ref{F0flat}) 
and (\ref{F2flat}), if we identify the flat space time $t$ with conformal time 
$\eta$. Because the de Sitter result (\ref{Sigma1}) must degenerate to its
flat space ancestor (\ref{flatrep}) for $H = 0$, we can extract terms 
proportional to the projectors (\ref{dS0}) and (\ref{dS2}), acting on 
$f_B(\Delta x)$,
\begin{eqnarray}
\lefteqn{ [\mbox{}^{\mu\nu} \Sigma^{\rho\sigma}](x;x') = \frac{-\kappa^2}{
2^{10} \!\cdot\! 3 \!\cdot\! \pi^3} \Bigl[ \tfrac13 \mathcal{F}^{\mu\nu} 
\!\!\times\! {\mathcal{F}'}^{\rho\sigma} \!\!+\! \tfrac15 \mathcal{C}_{\alpha
\beta\gamma\delta}^{~~~~~\mu\nu} \!\!\times\! {\mathcal{C}'}^{\alpha\beta\gamma
\delta\rho\sigma} \Bigr] \Bigl[ 8\pi \ln(a a') } \nonumber \\
& & \hspace{-0.5cm} \times \delta^4(\Delta x) \!+\! f_B(\Delta x) \Bigr] + 
\frac{\kappa^2 H^2}{2^7 \!\cdot\! 3 \!\cdot\! \pi^3} \sum_{i=1}^{21} 
\Bigl\{-2\pi \Delta T^i(a,a',\partial) [\mbox{}^{\mu\nu} D^{\rho\sigma}_{i}] 
\delta^4(\Delta x) \nonumber \\
& & \hspace{1cm} + \Delta T^i_{A}(a,a'\!,\partial) [\mbox{}^{\mu\nu} 
D^{\rho\sigma}_{i}] f_A (\Delta x) + \Delta T^i_{B}(a,a'\!,\partial) 
[\mbox{}^{\mu\nu} D^{\rho\sigma}_{i}] f_B(\Delta x) \Bigr\} . 
\label{Sigma2} \qquad
\end{eqnarray}
Table~\ref{Residual} gives the residual coefficient functions.
Note that the same combination of $8 \pi \ln(a a') \delta^4(\Delta x) + f_B(\Delta x)$ 
appears in both structure functions. This is because logarithms of $a a'$ and of
the dimensional regularization scale $\mu$, which appears in expression (\ref{fAfB})
for $f_B(\Delta x)$, arise from renormalization. For primitive diagrams it turns
out that $D$-dependent factors of $a$ are always canceled by inverse factors from
propagators, so primitive divergences contain no $D$-dependent factors of $a$
\cite{Miao:2024atw}. On the other hand, the counterterms which subtract these 
divergences all inherit a factor of $a^D$ from the $\sqrt{-g}$ in the measure.
Hence, there is an incomplete cancellation between primitive divergences and
counterterms,
\begin{equation}
\left(\mathrm{Primitive : }\frac{(\frac{H}{2})^{D-4}}{D \!-\! 4} \right) - 
\left(\mathrm{Counterterm: } \frac{(\mu a)^{D-4}}{D \!-\! 4} \right) = -
\ln\Bigl(\frac{ 2 \mu a}{H}\Bigr) + O(D \!-\! 4) \; . \label{origin}
\end{equation}
When all the propagators are doubly-differentiated (as in this calculation) this 
cancellation is the only source for factors of $\ln(a)$ and $\ln(\mu)$, which is
why they come in the combination $8\pi \ln(a a') \delta^4(\Delta x) + f_B(\Delta x)$.

\section{The Lichnerowicz Operator as a Bi-Tensor}

The purpose of this section is to extract from the summed terms of expression
(\ref{Sigma2}) a 3rd contribution proportional to the combination $8\pi \ln(a a')
\delta^4(\Delta x) + f_B(\Delta x)$. This is done by redefining the Lichnerowicz 
operator of equation (\ref{Einstein}) as a bi-tensor $[\mbox{}^{\mu\nu} 
\mathcal{L}^{\rho\sigma}]$. The simplest choice permits us to represent the 
logarithm parts of the $\Delta T^i$ sum in (\ref{Sigma2}) as proportional to 
$[\mbox{}^{\mu\nu} \mathcal{L}^{\rho\sigma}] \times 8\pi \delta^4(\Delta x)$. 
We then add factors proportional to $\Delta \eta$ to the $\Delta 
T^i_{B}(a,a',\partial)$ coefficient functions of Table~\ref{Residual}, and use 
the final relation of (\ref{relations}), to express the $\Delta T^i_{B}$ sum in 
(\ref{Sigma2}) as proportional to $[\mbox{}^{\mu\nu} \mathcal{L}^{\rho\sigma}] 
\times f_B(\Delta)$, plus a sum of terms proportional to $f_A(\Delta x)$. The 
section ends by noting that $\mathcal{W}^{\mu}_{~\alpha\beta} \times [\mbox{}^{\alpha\beta} 
\mathcal{L}^{\rho\sigma}] \neq 0$, so that the new contribution is not conserved 
without the $\delta T^i$ and $\delta T^i_{A}$ residues. 

On de Sitter background the Lichnerowicz operator is,
\begin{eqnarray}
\lefteqn{ \mathcal{D}^{\mu\nu\rho\sigma} h_{\rho\sigma} = \tfrac12 a^2 \Bigl[
\partial^2 h^{\mu\nu} - \eta^{\mu\nu} \partial^2 h + \eta^{\mu\nu} \partial^{\rho}
\partial^{\sigma} h_{\rho\sigma} + \partial^{\mu} \partial^{\nu} h - 2 
\partial^{\rho} \partial^{(\mu} h^{\nu)}_{~~\rho}\Bigr] } \nonumber \\
& & \hspace{0.5cm} + a^3 H \Bigl[ \eta^{\mu\nu} \partial_0 h - \partial_0 h^{\mu\nu}
- 2 \eta^{\mu\nu} \partial^{\rho} h_{\rho 0} + 2 \partial^{(\mu} h^{\nu)}_{~~0}
\Bigr] + 3 a^4 H^2 \eta^{\mu\nu} h_{00} \; . \qquad \label{Lichop}
\end{eqnarray}
We seek to define a bi-tensor operator $[\mbox{}^{\mu\nu} 
\mathcal{L}^{\rho\sigma}]$ such that,
\begin{equation}
\int \!\! d^4x' \, \Bigl\{ [\mbox{}^{\mu\nu} \mathcal{L}^{\rho\sigma}] 
\!\times\! \delta^4(\Delta x)\Bigr\} \!\times\! h_{\rho\sigma}(x') = 
\mathcal{D}^{\mu\nu\sigma\rho} h_{\rho\sigma}(x) \; . \label{desired}
\end{equation}
A number of choices are possible but the simplest is,
\begin{equation}
[\mbox{}^{\mu\nu} \mathcal{L}^{\rho\sigma}] \equiv \Bigl\{ [\eta^{\mu\nu} 
\eta^{\rho\sigma} \!-\! \eta^{\mu (\rho} \eta^{\sigma) \nu}] \partial 
\!\cdot\! \partial' + 2 {\partial'}^{(\mu} \eta^{\nu) (\rho} \partial^{\sigma)}
+ {\partial'}^{\mu} {\partial'}^{\nu} \eta^{\rho\sigma} + \eta^{\mu\nu}
\partial^{\rho} \partial^{\sigma} \Bigr\} \tfrac{a a'}{2} \; . \label{Lichbiten}
\end{equation}

\begin{table}[]
\setlength{\tabcolsep}{8pt}
\def\arraystretch{1.5}
\centering
\begin{tabular}{|@{\hskip 1mm }c@{\hskip 1mm }||c|c|c|}
\hline
$i$ & $\delta T^i(a,a',\partial)$ & $\delta T^i_{A}(a,a',\partial)$ &
$[\mbox{}^{\mu\nu} D_i^{\rho\sigma}]$ \\
\hline\hline
1 & $-3 a a' \partial^2 \!-\! 4 a^2 {a'}^2 H^2$ & 
$-\tfrac12 a^2 {a'}^2 H^2 \partial_0^2 \partial^2$ & $\eta^{\mu\nu} 
\eta^{\rho\sigma}$ \\
\hline
2 & $a a' \partial^2$ & $\tfrac12 a^2 {a'}^2 H^2 \partial_0^2 \partial^2$ &
$\eta^{\mu (\rho} \eta^{\sigma) \nu}$ \\
\hline
3 & $6 a^3 a' H^2$ & $-a^3 {a'}^2 H^3 \partial_0 \partial^2$ & 
$\eta^{\mu\nu} \delta^{\rho}_{~0} \delta^{\sigma}_{~0}$ \\
\hline
4 & $6 a {a'}^3 H^2$ & $a^2 {a'}^3 H^3 \partial_0 \partial^2$ &
$\delta^{\mu}_{~0} \delta^{\nu}_{~0} \eta^{\rho\sigma}$ \\
\hline
5 & $-6 a^2 a' H$ & $a^2 {a'}^2 H^2 \partial_0 \partial^2$ & $\eta^{\mu\nu}
\delta^{(\rho}_{~~0} \partial^{\sigma)}$ \\
\hline
6 & $6 a {a'}^2 H$ & $a^2 {a'}^2 H^2 \partial_0 \partial^2$ & $\delta^{(\mu}_{~~0}
\partial^{\nu)} \eta^{\rho\sigma}$ \\
\hline
7 & $3 a a'$ & $\tfrac12 a^2 {a'}^2 H^2 \partial^2$ & $\eta^{\mu\nu}
\partial^{\rho} \partial^{\sigma}$ \\
\hline
8 & $3 a a'$ & $\tfrac12 a^2 {a'}^2 H^2 \partial^2\!\!\!$ & $\partial^{\mu} 
\partial^{\nu} \eta^{\rho\sigma}$ \\
\hline
9 & $-2 a^2 {a'}^2 H^2$ & $0$ & $\delta^{(\mu}_{~~0} \eta^{\nu) (\rho} 
\delta^{\sigma)}_{~~0}$ \\
\hline
10 & $-2 a {a'}^2 H$ & $-a^2 {a'}^2 H^2 \partial_0 \partial^2$ & 
$\delta^{(\mu}_{~~0} \eta^{\nu) (\rho} \partial^{\sigma)}$ \\
\hline
11 & $2 a^2 a' H$ & $-a^2 {a'}^2 H^2 \partial_0 \partial^2$ & $\partial^{(\mu}
\eta^{\nu) (\rho} \delta^{\sigma)}_{~~0}$ \\
\hline
12 & $-2 a a'$ & $-a^2 {a'}^2 H^2 \partial^2$ & $\partial^{(\mu}
\eta^{\nu) (\rho} \partial^{\sigma)}$ \\
\hline
18 & $0$ & $\!\!\!-3 a^2 {a'}^2 H^2 \partial^2$ & $\delta^{(\mu}_{~~0} 
\partial^{\nu)} \delta^{(\rho}_{~~0} \partial^{\sigma)}$ \\
\hline
19 & $0$ & $\!\!\!a^2 a' H \partial^2\!\!\!$ & $\delta^{(\mu}_{~~0} 
\partial^{\nu)} \partial^{\rho} \partial^{\sigma}$ \\
\hline
20 & $0$ & $\!\!\!-a {a'}^2 H \partial^2\!\!\!$ & $\partial^{\mu}
\partial^{\nu} \delta^{(\rho}_{~~0} \partial^{\sigma)}$ \\
\hline
21 & $0$ & $\!\!\!\tfrac12 a a' \partial^2 \!-\! a^2 {a'}^2 H^2\!\!\!$ &
$\partial^{\mu} \partial^{\nu} \partial^{\rho} \partial^{\sigma}$ \\
\hline
\end{tabular}
\caption{\footnotesize Tabulation of the residues $\delta T^i(a,a',\partial)$
and $\delta T^i_{A}(a,a',\partial)$ which appear in expression (\ref{Sigma3}).}
\label{Residue}
\end{table}

\begin{eqnarray}
\lefteqn{ [\mbox{}^{\mu\nu} \Sigma^{\rho\sigma}](x;x') = \frac{-\kappa^2}{
2^{10} \!\cdot\! 3 \!\cdot\! \pi^3} \Bigl[ \tfrac13 \mathcal{F}^{\mu\nu} 
\!\!\times\! {\mathcal{F}'}^{\rho\sigma} \!\!+\! \tfrac15 \mathcal{C}_{\alpha
\beta\gamma\delta}^{~~~~~\mu\nu} \!\!\times\! {\mathcal{C}'}^{\alpha\beta\gamma
\delta\rho\sigma} \!\!+ 4 H^2 [\mbox{}^{\mu\nu} \mathcal{L}^{\rho\sigma}]
\Bigr] } \nonumber \\
& & \hspace{1cm} \times \Bigl[ 8\pi \ln(a a') \delta^4(\Delta x) \!+\! 
f_B(\Delta x) \Bigr] - \frac{\kappa^2 H^2}{2^7 \!\cdot\! 3 \!\cdot\! \pi^3} 
\sum_{i=1}^{21} \Bigl\{2\pi \delta T^i(a,a',\partial) \nonumber \\
& & \hspace{3.5cm} \times [\mbox{}^{\mu\nu} D^{\rho\sigma}_{i}] 
\delta^4(\Delta x) - \delta T^i_{A}(a,a',\partial) [\mbox{}^{\mu\nu} 
D^{\rho\sigma}_{i}] f_A (\Delta x) \Bigr\} . \label{Sigma3} \qquad
\end{eqnarray}

Unlike the Ricci and Weyl contributions to expression (\ref{Sigma3}),
the terms proportional to the Lichnerowicz operator are not independently
conserved. This is evident from the action of the Ward operator,
\begin{eqnarray}
\lefteqn{\mathcal{W}^{\mu}_{~\alpha\beta} \!\times\! [\mbox{}^{\alpha\beta} 
\mathcal{L}^{\rho\sigma}] = \delta^{\mu}_{~0} \Bigl\{\tfrac12 \eta^{\rho\sigma} 
\partial \!\cdot\! \partial' + \tfrac12 \partial^{\rho} \partial^{\sigma} } 
\nonumber \\
& & \hspace{3cm} + \tfrac12 \eta^{\rho\sigma} \partial'_0 a H \!+\! 
\delta^{(\rho}_{~~0} \partial^{\sigma)} a H \!+\! \delta^{\rho}_{~0} 
\delta^{\sigma}_{~0} \, a^2 H^2\Bigr\} a^2 {a'}^2 H^2 \Delta \eta \; . \qquad 
\label{WLop}
\end{eqnarray}
Neither the $\delta^4(\Delta x)$ term nor the $f_B(\Delta x)$ contribution
is independently conserved, nor is their sum. Conservation only follows when
one includes the sums over the ``residual'' terms involving $\delta T^i(a,a',
\partial)$ and $\delta T^i_{A}(a,a',\partial)$.

\section{Representing the Residue}

The purpose of this section is to recognize the remaining sum in expression
(\ref{Sigma3}) as differential operators acting on $\delta^4(\Delta x)$
and $f_A(\Delta x)$. We begin by summing the $\delta T^i$ terms to a relatively
compact form, then we do the same for the sum of the $T^i_{A}$ terms listed in Table 5. This
prompts the definition of two new differential operators,
$[\mbox{}^{\mu\nu} \mathcal{K}^{\rho\sigma}]$ and $[\mbox{}^{\mu\nu} 
\mathcal{J}^{\rho\sigma}]$, in terms of which the graviton self-energy is 
given its final form. The section closes with a consideration of conservation.

The $\delta T^i(a,a',\partial)$ terms can be represented as,
\begin{eqnarray}
\lefteqn{-2 \pi \sum_{i=1}^{21} \delta T^i(a,a',\partial) [\mbox{}^{\mu\nu} 
D_i^{\rho\sigma}] \delta^4(\Delta x) = -2\pi \Bigl\{ [3 \eta^{\mu\nu} 
\eta^{\rho\sigma} \!-\! \eta^{\mu (\rho} \eta^{\sigma) \nu} ] \partial \!\cdot\! 
\partial' } \nonumber \\
& & \hspace{3cm} + 3 [{\partial'}^{\mu} {\partial'}^{\nu} \eta^{\rho\sigma} 
\!+\! \eta^{\mu\nu} \partial^{\rho} \partial^{\sigma}] + 2 {\partial'}^{(\mu} 
\eta^{\nu) (\rho} \partial^{\sigma)} \Bigr\} \Bigl[a a' \delta^4(\Delta x)
\Bigr] \nonumber \\
& & \hspace{2cm} - 4\pi a^2 {a'}^2 H^2 \Bigl\{ \eta^{\mu\nu} \eta^{\rho\sigma}
\!-\! \eta^{\mu (\rho} \eta^{\sigma) \nu} \!-\! 2 \delta^{(\mu}_{~~0} 
\eta^{\nu) (\rho} \delta^{\sigma)}_{~~0} \Bigr\} \delta^4(\Delta x) \; , 
\qquad \\
& & \hspace{-0.5cm} = -4\pi [\mbox{}^{\mu\nu} \mathcal{L}^{\rho\sigma}]
\delta^4(\Delta x) - 4\pi \Bigl\{\eta^{\mu\nu} \eta^{\rho\sigma} \partial 
\!\cdot\! \partial' \!+\! {\partial'}^{\mu} {\partial'}^{\nu} \eta^{\rho\sigma}
\!+\! \eta^{\mu\nu} \partial^{\rho} \partial^{\sigma} \Bigr\} \Bigl[ a a' 
\delta^4(\Delta x)\Bigr] \nonumber \\
& & \hspace{2cm} - \tfrac12 a^2 {a'}^2 H^2 \Bigl\{ \eta^{\mu\nu} 
\eta^{\rho\sigma} \!-\! \eta^{\mu (\rho} \eta^{\sigma) \nu} \!-\! 2 
\delta^{(\mu}_{~~0} \eta^{\nu) (\rho} \delta^{\sigma)}_{~~0} \Bigr\}
\partial^4 f_A(\Delta x) \; . \qquad \label{deltaTrep}
\end{eqnarray}
When the three terms on the final line of (\ref{deltaTrep}) are added to
the sum over $\delta T^i_{A}$ we express the result as a new differential operator,
\begin{eqnarray}
\lefteqn{-\tfrac12 [\mbox{}^{\mu\nu} \mathcal{K}^{\rho\sigma}] f_A(\Delta x) 
= \sum_{i=1}^{21} \delta T^i(a,a',\partial) [\mbox{}^{\mu\nu} 
D_i^{\rho\sigma}] f_{A}(\Delta x) } \nonumber \\
& & \hspace{2cm} - \tfrac12 a^2 {a'}^2 H^2 \Bigl\{ \eta^{\mu\nu} 
\eta^{\rho\sigma} \!-\! \eta^{\mu (\rho} \eta^{\sigma) \nu} \!-\! 2 
\delta^{(\mu}_{~~0} \eta^{\nu) (\rho} \delta^{\sigma)}_{~~0} \Bigr\}
\partial^4 f_A(\Delta x) \; . \qquad
\label{deltaTArep}
\end{eqnarray}
This operator is,
\begin{eqnarray}
\lefteqn{ [\mbox{}^{\mu\nu} \mathcal{K}^{\rho\sigma}] \equiv a^2 {a'}^2 H^2
\Bigl\{ \Pi^{\mu\nu} \Pi^{\rho\sigma} \!\!-\! \Pi^{\mu (\rho} 
\Pi^{\sigma) \nu} \!\!+\! [\eta^{\mu\nu} \partial_0 \!-\! 2 \delta^{(\mu}_{~~0}
\partial^{\nu)} ] [\eta^{\rho\sigma} \partial_0 \!-\! 2 \delta^{(\rho}_{~~0}
\partial^{\sigma)} ] \partial^2 - } \nonumber \\
& & \hspace{-0.5cm} \Bigl[ \eta^{\mu (\rho} \eta^{\sigma) \nu} \partial_0^2
\!\!-\! 2 \delta^{(\mu}_{~0} \eta^{\nu) (\rho} \partial^{\sigma)} \partial_0
\!\!-\! 2 \partial^{(\mu} \eta^{\nu) (\rho} \delta^{\sigma)}_{~0} \partial_0
\!\!+\! 2 \delta^{(\mu}_{~0} \eta^{\nu) (\rho} \delta^{\sigma)}_{~0} 
\partial^2 \!\!+\! 2 \delta^{(\mu}_{~0} \partial^{\nu)} \delta^{(\rho}_{~0} 
\partial^{\sigma)} \Bigr] \partial^2 \nonumber \\
& & \hspace{1.5cm} - \partial^{\mu} \partial^{\nu} {\partial'}^{\rho} 
{\partial'}^{\sigma} \!\Bigl[ \frac{\partial^2}{a a' H^2} \!-\! 2 \Bigr] 
\!+\! 2 H [ a \eta^{\mu\nu} \delta^{\rho}_{~0} \delta^{\sigma}_{~0} \!-\!
a' \delta^{\mu}_{~0} \delta^{\nu}_{~0} \eta^{\rho\sigma}] \partial_0 \partial^2
\! \Bigr\} . \qquad \label{Kdef}
\end{eqnarray}
We can give the graviton self-energy a relatively compact form by defining
a final differential operator,
\begin{equation}
[\mbox{}^{\mu\nu} \mathcal{J}^{\rho\sigma}] \equiv \Bigl\{ \eta^{\mu\nu}
\eta^{\rho\sigma} \partial \!\cdot\! \partial' \!+\! {\partial'}^{\mu}
{\partial'}^{\nu} \eta^{\rho\sigma} \!+\! \eta^{\mu\nu} \partial^{\rho}
\partial^{\sigma} \Bigr\} a a' \; . \label{Jdef}
\end{equation}
The resulting form for $[\mbox{}^{\mu\nu} \Sigma^{\rho\sigma}](x;x')$ is,
\begin{eqnarray}
\lefteqn{ [\mbox{}^{\mu\nu} \Sigma^{\rho\sigma}] = \frac{-\kappa^2}{
2^{10} \!\cdot\! 3 \!\cdot\! \pi^3} \Bigl[ \tfrac13 \mathcal{F}^{\mu\nu} 
{\mathcal{F}'}^{\rho\sigma} \!\!\!+\! \tfrac15 \mathcal{C}_{\alpha
\beta\gamma\delta}^{~~~~~\mu\nu} {\mathcal{C}'}^{\alpha\beta\gamma
\delta\rho\sigma} \Bigr] \!\! \Bigl[ 8\pi \ln(a a') \delta^4(\Delta x) 
\!+\!\! f_B(\Delta x) \Bigr] } \nonumber \\
& & \hspace{2cm} - \frac{\kappa^2 H^2}{2^8 \!\cdot\! 3 \!\cdot\! \pi^3} 
\Bigl\{ [\mbox{}^{\mu\nu} \mathcal{L}^{\rho\sigma}] \Bigl[ 8\pi [\ln(a a') 
\!+\! 1] \delta^4(\Delta x) + f_B(\Delta x)\Bigr] \nonumber \\
& & \hspace{5cm} + [\mbox{}^{\mu\nu} \mathcal{J}^{\rho\sigma}] 8\pi 
\delta^4(\Delta x) + [\mbox{}^{\mu\nu} \mathcal{K}^{\rho\sigma}] 
f_A(\Delta x) \Bigr\} . \label{Sigma4} \qquad
\end{eqnarray}
All three structure functions can be read from (\ref{Sigma4}),
\begin{eqnarray}
F_1(x;x') &=& - \tfrac{\kappa^2}{2^{10} \cdot 3^2 \cdot \pi^3} 
\times \Bigl[ 8\pi \ln(a a') \delta^4(\Delta x) \!+\! f_B(\Delta x) \Bigr]  
\; , \qquad \label{sf1} \\
F_2(x;x') &=& - \tfrac{\kappa^2}{2^{10} \cdot 3 \cdot 5 \cdot \pi^3} 
\times \Bigl[ 8\pi \ln(a a') \delta^4(\Delta x) \!+\! f_B(\Delta x) \Bigr]  
\; , \qquad \label{sf2} \\
F_3(x;x') &=& - \tfrac{\kappa^2 H^2}{2^8 \cdot 3 \cdot \pi^3} 
\Bigl[ 8\pi \ln(a a') \delta^4(\Delta x) \!+\! f_B(\Delta x) \Bigr] 
\; . \qquad \label{sf3}
\end{eqnarray}
The third structure function is associated with two additional terms
needed for conservation,
\begin{equation}
F_{3a}(\Delta x) = -\tfrac{\kappa^2 H^2}{2^8 \cdot 3 \cdot \pi^3} \times
8 \pi \delta^4(\Delta x) \quad , \quad F_{3b}(\Delta x) = -
\tfrac{\kappa^2 H^2}{2^8 \cdot 3 \cdot \pi^3} \times f_A(\Delta x) \; .
\label{sf3ab}
\end{equation}
We might identify an additional, 4th structure function as,
\begin{equation}
F_4(\Delta x) = - \tfrac{\kappa^2 H^2}{2^8 \cdot 3 \cdot \pi^3}
\times 8 \pi \delta^4(\Delta x) \; . \label{sf4}
\end{equation}
The final representation is,
\begin{eqnarray}
\lefteqn{ [\mbox{}^{\mu\nu} \Sigma^{\rho\sigma}] = 
\mathcal{F}^{\mu\nu} {\mathcal{F}'}^{\rho\sigma} F_1
+ \mathcal{C}_{\alpha\beta\gamma\delta}^{~~~~~\mu\nu} 
{\mathcal{C}'}^{\alpha\beta\gamma\delta\rho\sigma} F_2 }
\nonumber \\
& & \hspace{2cm} + \Bigl\{[\mbox{}^{\mu\nu} \mathcal{L}^{\rho\sigma}] 
F_3 + [\mbox{}^{\mu\nu} \mathcal{J}^{\rho\sigma}] F_{3a} +
[\mbox{}^{\mu\nu} \mathcal{K}^{\rho\sigma}] F_{3b}\Bigr\}
+ [\mbox{}^{\mu\nu} \mathcal{L}^{\rho\sigma}] F_4 \; . 
\qquad \label{final}
\end{eqnarray}
Whereas the contributions on the first line of (\ref{final}) are 
conserved for any $F_1(x;x)$ and $F_2(x;x')$, the conservation of
the second line contributions depends on the functional form of 
the associated structure functions.
 
Although the combination of all the ``new'' terms on the last two lines of 
expression (\ref{Sigma4}) are conserved, the only separately conserved
constituent is $[\mbox{}^{\mu\nu} \mathcal{L}^{\rho\sigma}] 
\delta^4(\Delta x)$. Its conservation follows from relation (\ref{WLop}).
The action of the Ward operator on the $f_B$ term is,
\begin{eqnarray}
\lefteqn{ \mathcal{W}^{\mu}_{~\alpha\beta} \!\times\! [\mbox{}^{\alpha\beta} 
\mathcal{L}^{\rho\sigma}] f_{B}(\Delta x) = 8\pi a^2 {a'}^2 H^2 
\delta^{\mu}_{~0} \eta^{\rho\sigma} \partial_0 \delta^4(\Delta x) -
a^2 {a'}^2 H^2 \delta^{\mu}_{~0} \Bigl\{ \partial^{\rho}
\partial^{\sigma} } \nonumber \\
& & \hspace{-0.5cm} - 2 a H \delta^{(\rho}_{~~0} \partial^{\sigma)} \!\!+\!
2 a^2 H^2 \delta^{\rho}_{~0} \delta^{\sigma}_{~0} \!+\! \eta^{\rho\sigma} \! 
\Bigl[a H \partial_0 \!-\! 2 a' H \partial_0 \!-\! 2 a a' H^2 \Bigr] \!
\Bigr\} \partial_0 \partial^2 \! f_A(\Delta x) \, . \quad \label{WLB}
\end{eqnarray}
This is very largely canceled by the $[\mbox{}^{\mu\nu} 
\mathcal{K}^{\rho\sigma}] f_A(\Delta x)$ term,
\begin{eqnarray}
\lefteqn{ \mathcal{W}^{\mu}_{~\alpha\beta} \!\!\times\!\! [\mbox{}^{\alpha\beta} 
\mathcal{K}^{\rho\sigma}] f_{A}(\Delta x) = -8\pi a^2 {a'}^2 \! \Bigl\{ \! 
H^2 \delta^{\mu}_{~0} \eta^{\rho\sigma} \partial_0 \!\!+\! (\partial^{\mu} 
\!+\! a H \delta^{\mu}_{~0}) {\partial'}^{\rho} \! {\partial'}^{\sigma} 
\!\! \frac1{a a'} \!\! \Bigr\} \delta^4(\Delta x) } \nonumber \\
& & \hspace{2cm} + a^2 {a'}^2 H^2 \delta^{\mu}_{~0} \Bigl\{ \partial^{\rho}
\partial^{\sigma} \!-\! 2 a H \delta^{(\rho}_{~~0} \partial^{\sigma)} \!+\!
2 a^2 H^2 \delta^{\rho}_{~0} \delta^{\sigma}_{~0} \nonumber \\
& & \hspace{4cm} + \eta^{\rho\sigma} \Bigl[a H \partial_0 \!-\! 2 a' H 
\partial_0 \!-\! 2 a a' H^2\Bigr] \Bigr\} \partial_0 \partial^2 f_A(\Delta x) 
. \qquad \label{WK}
\end{eqnarray}
The remaining contributions are proportional to $\delta^4(\Delta x)$,
\begin{eqnarray}
\lefteqn{ \mathcal{W}^{\mu}_{~\alpha\beta} \!\times\! [\mbox{}^{\alpha\beta} 
\mathcal{J}^{\rho\sigma}] \Bigl[8\pi \delta^4(\Delta x)\Bigr] = 8\pi
(\partial^{\mu} \!+\! 4 a H \delta^{\mu}_{~0}) \partial^{\rho}
\partial^{\sigma} \Bigl[a a' \delta^4(\Delta x)\Bigr] } \nonumber \\
& & \hspace{1cm} + 8\pi H \delta^{\mu}_{~0} \eta^{\rho\sigma} \partial 
\!\cdot\! \partial' \Bigl[a {a'}^2 \delta^4(\Delta x)\Bigr] + 8\pi H^2
\delta^{\mu}_{~0} \eta^{\rho\sigma} \partial_0' \Bigl[a^2 {a'}^2 
\delta^4(\Delta x)\Bigr] \; , \qquad \label{WJ} \\
\lefteqn{ \mathcal{W}^{\mu}_{~\alpha\beta} \!\times\! [\mbox{}^{\alpha\beta} 
\mathcal{L}^{\rho\sigma}] \Bigl[8\pi \ln(a a') \delta^4(\Delta x)\Bigr]
= -8\pi H \delta^{\mu}_{~0} (\eta^{\rho\sigma} \partial \!\cdot\! \partial'
\!+\! \partial^{\rho} \partial^{\sigma}) [a^3 \delta^4(\Delta x)] } 
\nonumber \\
& & \hspace{0.4cm} - 8 \pi H^2 \delta^{\mu}_{~0} \Bigl[\eta^{\rho\sigma} 
\partial'_0 \!+\! 2 \delta^{(\rho}_{~~0} \partial^{\sigma)} \Bigr] [a^4 
\delta^4(\Delta x)] \!-\! 16 \pi H^3 \delta^{\mu}_{~0} \delta^{\rho}_{~0} 
\delta^{\sigma}_{~0} a^5 \delta^4(\Delta x) \; . \qquad \label{WLlog}
\end{eqnarray}

\section{Epilogue}

On flat space background, matter contributions to the graviton self-energy
take the form (\ref{flatrep}) of two 4th order differential operators 
acting on scalar structure functions. One of the 4th order operators has
spin zero and the other has spin two, and each of them is transverse, so 
that each contribution is conserved, no matter what the form of the scalar 
structure functions. On de Sitter background (\ref{deSitter}) these operators
generalize to (\ref{dS0}) and (\ref{dS2}), respectively. They are based on 
differential operators (\ref{Fdef}) and (\ref{Cdef}), which are 
constructed from the linearized Ricci and Weyl tensors, respectively. These
operators are each annihilated by the Ward operator (\ref{Wardop}), so their
contributions are again conserved, no matter what structure function they
act upon. These contributions to the graviton self-energy are given in
expression (\ref{Sigma2}), whose scalar structure functions agree with 
those of flat space (\ref{F0flat}) and (\ref{F2flat}), except for factors
of $8\pi \ln(a a') \delta^4(\Delta x)$, which derive from renormalization
(\ref{origin}).

Our concern in this paper is with the new contributions which arise from the
cosmological background. They appear in expression (\ref{Sigma2}) as sums
of functions of the two scale factors and the derivative operator, given in
Table~\ref{Residual}, multiplying the 21 differential operators given 
in Table~\ref{Tbasis}. Our goal has been to represent them as a 2nd order 
differential operator acting on $\delta^4(\Delta x)$ and the functions
$f_A(\Delta x)$ and $f_B(\Delta x)$ defined in expression (\ref{fAfB}). The
alternative is to represent the new terms using a 4th order differential
operator constructed from contracting two factors of (\ref{Cdef}) with 
constant index factors such as the spatial Minkowski metric (\ref{spaceeta}). The
result is very complicated structure functions \cite{Leonard:2014zua}, which
are difficult to use in solving the effective field equations 
\cite{Park:2015kua}.

Our final form is expression (\ref{Sigma4}), which is based on the 
Lichnerowicz bi-tensor $[\mbox{}^{\mu\nu} \mathcal{L}^{\rho\sigma}]$, given 
in (\ref{Lichbiten}). However, two additional differential operators
are required: $[\mbox{}^{\mu\nu} \mathcal{K}^{\rho\sigma}]$, given in
(\ref{Kdef}), and $[\mbox{}^{\mu\nu} \mathcal{J}^{\rho\sigma}]$, given in
(\ref{Jdef}). Each of these differential operators acts on a different
combination of $\delta^4(\Delta x)$, $f_A(\Delta x)$ and $f_B(\Delta x)$,
and relations (\ref{WLB}-\ref{WLlog}) show that they must be combined 
so as to produce a conserved result. 

Consideration of Table~\ref{Residual} suggests that we consider moving
all the $\partial_0 \partial^2$ contributions in the $\Delta T^i_{A}$
column to the $\Delta T^i_{B}$ column using relation (\ref{relations}).
One would then define a new Lichnerowicz bi-tensor to preserve the form
$[\mbox{}^{\mu\nu} \mathcal{L}^{\rho\sigma}] [8\pi \ln(a a') 
\delta^4(\Delta x) + f_B(\Delta x)]$. There would be corresponding 
changes in the residual $\delta^4(\Delta x)$ and $f_A(\Delta x)$ 
contributions. We explore this possibility in the Appendix. The choice 
seems to be between having a simple Lichnerowicz bi-tensor, with a complicated
$[\mbox{}^{\mu\nu} \mathcal{K}^{\rho\sigma}]$, or the converse.

This form (\ref{Sigma4}), with different multiples of the three contributions,
must pertain for any matter contribution to the graviton self-energy on de
Sitter background. For example, the conformal invariance of $D=4$ electrodynamics
restricts the contribution from a single loop of photons to just the Weyl term
\cite{Wang:2015eaa,Foraci:2024vng},
\begin{equation}
[\mbox{}^{\mu\nu} \Sigma^{\rho\sigma}_{\rm EM}](x;x') = -\frac{\kappa^2}{
2^8 \!\cdot\! 5 \!\cdot\! \pi^3} \, \mathcal{C}_{\alpha\beta\gamma\delta}^{
~~~~\mu\nu} \!\times\! {\mathcal{C}'}^{\alpha\beta\gamma\delta \rho\sigma}
\Bigl[ 8\pi \ln(a a') \delta^4(\Delta x) \!+\! f_B(\Delta x)\Bigr] .
\label{SigmaEM}
\end{equation}
The results for a loop of massless, Dirac fermions, and for a
loop of massless, conformally coupled scalars, take the same form as
(\ref{SigmaEM}) with factors of $\tfrac12$ and $\tfrac1{12}$, respectively
\cite{Foraci:2024cwi}. The relative factors are inherited from old flat space
results \cite{Capper:1973mv,Capper:1973bk,Capper:1974ed,Duff:2000mt}, for 
which the scale factor is $a=1$. It would be interesting to work out the 
contributions from massive fermions and for scalars with arbitrary mass and
conformal coupling. Also, it should be noted that the 3rd structure function 
is bound to occur in the graviton self-energy from a loop of gravitons, 
although there will be additional contributions as well. 

\appendix
\section{Appendix}
\setcounter{equation}{0}
\renewcommand{\theequation}{\thesection\arabic{equation}}

The purpose of this appendix is to explore the consequences of using the
identity $\partial_0 \partial^2 f_A(\Delta x) = -\tfrac12 \Delta \eta f_B(\Delta x)$ 
to simplify the $\Delta T^i_{A}(a,a',\partial)$ column of Table~\ref{Residual}.
If we want the modified $\Delta T^i_{B}(a,a',\partial)$ terms to still sum 
to ``the Lichnerowicz operator'' acting on $f_B(\Delta x)$ this obviously entails 
a redefinition of $[\mbox{}^{\mu\nu} \mathcal{L}^{\rho\sigma}]$,
\begin{eqnarray}
\lefteqn{ \left[\mbox{}^{\mu \nu}\!\mathcal{L}^{\rho \sigma}_{\rm new}\right] = 
\frac{aa'}{2} \Bigl\{\left(\eta^{\mu\nu}\eta^{\rho \sigma} \!-\! \eta^{\mu ( \rho} 
\eta^{\sigma ) \nu} \right) \left[\partial \!\cdot\! \partial' \!-\! 2 a a' H^2 
\right] \!+\! \eta^{\mu\nu} [\partial^{\rho} \partial^{\sigma} \!-\! 2 a' H 
\delta^{(\rho}_{~0} \partial^{\sigma)} ] } \nonumber \\
& & \hspace{1cm} + \eta^{\rho \sigma} [\partial^{\prime \mu} 
\partial^{\prime \nu} \!-\! 2 a H \delta^{(\mu}_{~~0} \partial^{\prime \nu)} ] 
+ 2 \Bigl[\partial^{\prime (\mu} \eta^{\nu)(\rho} \partial^{\sigma)} \!-\! a H 
\delta^{(\mu}_{~~0} \eta^{\nu)(\rho} \partial^{\sigma)} \nonumber \\
& & \hspace{4.8cm} - a' H \partial^{\prime (\mu} \eta^{\nu)(\rho} 
\delta^{\sigma)}_{~~0} \!+\! 3 a a' H^2 \delta^{(\mu}_{~~0} \eta^{\nu)(\rho} 
\delta^{\sigma)}_{~~0} \Bigr] \Bigr\} . \qquad \label{Lnew}
\end{eqnarray}
There is no change in $[\mbox{}^{\mu\nu} \mathcal{J}^{\rho\sigma}]$ from (\ref{Jdef}),
but $[\mbox{}^{\mu\nu} \mathcal{K}^{\rho\sigma}]$ is simplified to,
\begin{eqnarray}
\lefteqn{ [\mbox{}^{\mu \nu} \mathcal{K}^{\rho \sigma}_{\rm new}] = a^2 {a'}^2 H^2 
\Bigl\{ \Pi^{\mu \nu} \Pi^{\rho \sigma} \!-\! \Pi^{\mu (\rho}\Pi^{\sigma) \nu} \!-\!
\partial^{\mu} \partial^{\nu} {\partial'}^{\rho} {\partial'}^{\sigma} \Bigl( 
\tfrac{\partial^2}{a a' H^2} \!-\! 2 \Bigr) } \nonumber \\
& & \hspace{1.9cm} - 2 \delta^{(\mu}_{~~0} \partial^{\nu)} \delta^{(\rho}_{~~0}
\partial^{\sigma)} \partial^2 \!-\! [\eta^{\mu \nu} \eta^{\rho \sigma} \!-\! 
\eta^{\mu (\rho} \eta^{\sigma) \nu} \!+\! 2 \delta^{(\mu}_{~~0} \eta^{\nu)(\rho}
\delta^{\sigma)}_{~~0} ] \partial^4 \Bigr\} . \qquad \label{newK}
\end{eqnarray}
The full result (\ref{Sigma4}) then preserves its form.

We conclude this appendix by showing how conservation works with the new 
operators. Because $[\mbox{}^{\mu\nu} \mathcal{J}^{\rho\sigma}]$ is 
unaffected there is of course no change in relation (\ref{WJ}). The analog 
of (\ref{WLB}) is ,
\begin{eqnarray}
\lefteqn{ \mathcal{W}^{\mu}_{~\alpha \beta} \!\times\! [\mbox{}^{\alpha \beta} 
\mathcal{L}^{\rho \sigma}_{\rm new}] f_B(\Delta x) = 8 \pi a^2 {a'}^2 H^2 
\eta^{\mu(\rho} \delta^{\sigma)}_{~~0} \partial_0 \delta^4(\Delta x) } 
\nonumber \\
& & \hspace{1cm} - 2 a^2 {a'}^2 H^2 \Bigl\{ \tfrac12 \partial^{\mu} 
\delta^{(\rho}_{~~0} \partial^{\sigma)} \!+\! a H \partial^{\mu} 
\eta^{\rho \sigma} \!-\! a H \eta^{\mu( \rho} \partial^{\sigma)} \Bigr\} 
\partial_0 \partial^2 f_A(\Delta x) \; . \qquad 
\end{eqnarray}
The $f_A(\Delta x)$ terms of this are canceled by the analog of (\ref{WK}), 
\begin{eqnarray}
\lefteqn{ \mathcal{W}^{\mu}_{~\alpha \beta} \!\times\! [\mbox{}^{\mu \nu} 
\mathcal{K}^{\rho \sigma}_{\rm new}] f_A(\Delta x) = a^2 {a'}^2 H^2 \Bigl\{
\partial^{\mu} \delta^{(\rho}_{~~0} \partial^{\sigma)} \!+\! 2 a H \partial^{\mu}
\eta^{\rho \sigma} \!-\! 2 a H \eta^{\mu( \rho} \partial^{\sigma)} \Bigr\} }
\nonumber \\
& & \hspace{-0.7cm} \times \partial_0 \partial^2 f_A(\Delta x) - 8 \pi a a'
\Bigl\{(\partial^{\mu} \!+\! 2 a H \delta^{\mu}_{~0}) \partial^{\rho} 
\partial^{\sigma} \!-\! a' H (\partial^{\mu} \!+\! 2 a H \delta^{\mu}_{~0})
\delta^{(\rho}_{~~0} \partial^{\sigma)} \nonumber \\
& & \hspace{-0.7cm} + a a' H^2 (\partial^{\mu} \!\!+\! a H \delta^{\mu}_{~0}) 
\eta^{\rho \sigma} \!\!\!-\! a a' H^2 \eta^{\mu (\rho} \partial^{\sigma)} \!\!+\! 
a a' H^2 (\partial_0 \!+\! 2 a H) \eta^{\mu (\rho} \delta^{\rho)}_{~~0} \!\Bigr\} 
\delta^4(\Delta x) . \qquad 
\end{eqnarray}
The $\delta^4(\Delta x)$ terms are canceled by (\ref{WJ}) and the analog of
(\ref{WLlog}),
\begin{eqnarray}
\lefteqn{ \mathcal{W}^{\mu}_{~\alpha \beta} \!\!\times\!\! [\mbox{}^{\alpha \beta}\!
\mathcal{L}^{\rho \sigma}_{\rm new}] \! \Bigl[8 \pi \ln(a a') \delta^4(\Delta x)
\!\Bigr] \!\!=\! 8 \pi a a' H \Bigl\{ \! a a' H \partial^{\mu} \eta^{\rho \sigma} 
\!\!\!-\! a a' H \eta^{\mu (\rho} \partial^{\sigma)} \!\!-\! a \delta^{\mu}_{0} 
\partial^{\rho} \partial^{\sigma} } \nonumber \\
& & \hspace{0.5cm} + 2 a^2 a' H^2 \eta^{\mu (\rho} \delta^{\sigma)}_{~0} \!\!+\! 2 a a' H
\delta^{\mu}_{~0} \delta^{(\rho}_{~~0} \partial^{\sigma)} \!\!+\! \delta^{\mu}_{~0}
\eta^{\rho \sigma} a' (\partial^2 \!\!+\! 3 a a' H^2 ) \!\Bigr\} \delta^4(\Delta x) 
. \qquad
\end{eqnarray}
\vskip 0.5cm

\centerline{\bf Acknowledgements}

This work was partially supported by Taiwan NSTC grants 
112-2112-M-006-017 and 113-2112-M-006-013, by NSF grant 
PHY-2207514 and by the Institute for Fundamental Theory 
at the University of Florida.

\end{document}